\documentstyle[12pt]{article}

\newcommand{\be}{\begin{equation}}
\newcommand{\ee}{\end{equation}}
\newcommand{\bea}{\begin{eqnarray}}
\newcommand{\eea}{\end{eqnarray}}
\newcommand{\r}{\rho}
\newcommand{\th}{\varphi}
\newcommand{\rai}{\rightarrow \infty}

\def\lambdabar{{\mathchar'26\mkern-9mu\lambda}}
\def\cwl{\lambdabar_c}
\def\lb#1{\label{eq:#1}}
\def\rf#1{(\ref{eq:#1})}

\newcommand{\E}{{\cal E}_0}
\def\sp#1{\langle #1 \rangle }
\newcommand{\rb}{{\bf r}}
\newcommand{\rbh}{\hat{\bf r}}

\begin{document}

\baselineskip=14pt plus 0.2pt minus 0.2pt
\lineskip=14pt plus 0.2pt minus 0.2pt

\begin{flushright}
 hepth@xxx/9408057 \\
 LA-UR-94-2075 \\
\end{flushright}

\begin{center}
\Large{\bf
 EXACT, ${\bf E=0}$,  SOLUTIONS FOR GENERAL
POWER-LAW  POTENTIALS. \\
 I. CLASSICAL ORBITS} \\

\vspace{0.25in}

\large

\bigskip

Jamil Daboul\footnote{Email:  daboul@bguvms.bgu.ac.il}\\
{\it Physics Department, Ben Gurion University of the Negev\\
Beer Sheva, Israel}\\
$~~~~~~~$\\
and\\
$~~~~~~~$ \\
Michael Martin Nieto\footnote{Email:  mmn@pion.lanl.gov}\\
{\it
Theoretical Division, Los Alamos National Laboratory\\
University of California\\
Los Alamos, New Mexico 87545, U.S.A. }

\normalsize

\vspace{0.3in}

{ABSTRACT}

\end{center}

%********************************************************************
%\baselineskip=.33in
%*********************************************************************

\begin{quotation}
For zero energy, $E=0$,
we derive exact, classical solutions for {\em all}
power-law
potentials, $V(r)=-\gamma/r^\nu$,
with $\gamma>0$ and $-\infty <\nu<\infty$.
When the angular momentum is non-zero,
these solutions lead to the orbits  $\r(t)=
 [\cos \mu (\th(t)-\th_0(t))]^{1/\mu}$,
for all $\mu \equiv \nu/2-1 \ne 0$.
When $\nu>2$, the orbits are bound and go through the origin.
This leads to discrete discontinuities in the functional
dependence of $\th(t)$ and $\th_0(t)$, as functions of $t$, as the orbits
pass through the origin. We  describe a procedure
to connect  different analytic solutions for successive orbits at
the origin.
We calculate the periods and precessions of these bound orbits, and
graph a number of specific examples.  Also, we
explain why they all must violate the
virial theorem.  The unbound orbits are also discussed in detail.
This includes the unusual orbits
which have finite travel times
to infinity and also the special $\nu = 2$ case.
\vspace{0.25in}


\end{quotation}

\vspace{0.3in}

\newpage

\section{Introduction}

Since the birth of quantum mechanics, studying the connections between
classical
and quantum physics has been an enormous industry \cite{QM1,QM2,QM3}.  Both
the  intuitive and also the analytic aspects have been studied.  In
particular, one can observe a kind of ``Folk Theorem." Problems that are
amenable to well-defined,  exact, analytic solutions tend to be solvable in
both the classical
and the quantum regimes.

For example, the
fundamental three-dimensional problems of quantum mechanics, the
harmonic oscillator and the hydrogen atom, are those classical problems
which, by Bertrand's theorem \cite{bertrand},
 have exact, closed orbits for all energies.   The standard,
one-dimensional, quantum-mechanical potential problems that are exactly
solvable,
such as the Morse \cite{morse}, Rosen-Morse \cite{rm}, and P\"oschl-Teller
\cite{pt} potentials, are also solvable classically \cite{morsec,rmc,ptc}.
Further, it has even been shown that  quantum problems which are solvable,
but have certain ``mathematical diseases," manifest these diseases already
in the classical problem \cite{klauder}.

Usually,
 when one solves a potential system, one takes a particular potential
from a family and solves it for all values of the energy, $E$.   Here,
we are going to do the opposite.  We will
consider the entire class of
power-law potentials, parametrized as
\be
V(r) = -~\frac{\gamma}{r^{\nu}} = -\frac{\gamma}{r^{2\mu+2}}~,
{}~~~~~~\gamma > 0~,
 ~~~~~~  - \infty <\nu<\infty~, \lb{pot}
\ee
 and exactly solve it
 for all $\nu$
with the particular energy $E=0$.  Note that
it  will  be useful to switch back and forth between the variables
$\nu$ and $\mu$, related by
\be
\mu = (\nu-2)/2, ~~~~~~~ \nu = 2(\mu+1).
\ee
The potentials \rf{pot} are attractive for $\nu>0$ and repulsive for
$\nu <0$. For $\nu=0$,  the potential
is a constant, $V(r)=-\gamma$, so the particle is force-free.

The above system is exactly solvable in both the classical and
quantum-mechanical cases, and  there are similarities in the properties
of the solutions.
In the present paper, hereafter called I, we shall
 discuss the very unusual and enlightening  classical
solutions for the orbits. In the following paper  \cite{dn2}, called II,
we will solve the quantum problem and discuss the wave functions.

In Appendix A we give, for completeness, the simple solutions for zero
angular momentum.  The main body of the paper concentrates on the
more interesting general solutions, with non-zero, angular momentum.

We begin, in Sec. 2, by defining our ``dynamical" units,
which allow us to simplify the equations and their solutions.
 Then we  show that for $\nu \neq 2$, the classical solutions
have only one finite, non-zero turning radius, which we
will denote by $a$. This
$a$ has opposite interpretations, depending on the value of the power
index $\nu$.  For $\nu > 2$ the solutions yield bound orbits with
 $0 \leq r \leq a$, but for $\nu < 2$ they yield  unbound orbits,
 with $r \geq a$.
In Sec. 3 we  derive the general $E=0$ solutions, using two methods.
In Sec. 4 we investigate the general properties of the bound orbits,
$\nu>2$ or $\mu >0 $, such as their precessions and periods.

Continuing, in Sec. 5 we  graphically
depict  examples of the bound orbits, thereby demonstrating
the general properties derived previously.
Beginning with the case
$\nu=4$, which is a boundary case that does not precess,
we find that the  orbit  describes a circle
which goes through the origin and continually repeats itself.  For $\nu>4$,
the typical solution looks like a flower.  A single orbit starts
at the origin, goes out to $a$ and returns to the origin.  This first
orbit describes  what we will call a  ``petal." The following orbits
then go on to
describe further petals which in general are precessed from each other.
  After a number of petals, the trajectory closes
if $\nu$
is a rational fraction.  The petals become narrower and
narrower as $\nu \rightarrow \infty$.
Contrariwise,  for $4 > \nu > 2$, the orbits become tighter and
tighter spirals in and out of the origin as $\nu \rightarrow 2$.
 Once again, for $\nu$ being a rational fraction,
the orbits  eventually close on themselves.

In Secs. 6 and 7 we discuss the classically unbound orbits.  These
are given by
all $\nu<2$ as well as the special case $\nu = 2$.
We close with comments on the classical problem
and defer comparisons between the classical and quantum
problems to paper II.  Appendix B   discusses the scaling properties of
the solutions.  Appendix C investigates
 violations of the virial theorem.

%***********************************************************

\section{Classical Units and  $E=0$ Orbits}

Power-law potentials  do not have a built-in length
scale. To specify a power potential completely, two units
are needed:
a unit of energy, $\E$, and a unit of length, $a$. Instead of the unit of
energy, one can use a unit of angular momentum, $L_0$, which can be
related to $\E$, for example, by
\be
 \E=L_0^2/2ma^2~.
\ee
 Thus,
we can parametrize the power potentials as follows:
\be
  V(r) \equiv - \frac{\gamma}{r^\nu}
        \equiv - \E  \frac{g^2}{\r^\nu}
= - \frac{L^2_0}{2 m a^2} \frac{g^2}{\r^\nu}~, ~~~~
\r \equiv r/a~.       \lb{potN}
\ee
Therefore, the dimensional coupling constant, $\gamma$, is related to
the dimensionless coupling constant, $g^2$, by
\be
\gamma = \frac{L^2_0 a^{\nu-2}}{2 m }\;  g^2 ~.   \lb{Dgamma}
\ee

Dimensionless units, such as $g^2$, are convenient in quantum mechanics.
There one
has other physical constants, such as Planck's constant,
$\hbar$, to make a convenient choice of
 $L_0$ and  $\E$.  For example, in  Sec. 2 of II we choose
$L_0=\hbar$ to obtain $ \E=\hbar^2/2ma^2$ as the unit of energy, but we
leave $a$ arbitrary.
(We also could have  chosen
$a$ to be equal to
$\hbar/mc=\cwl$, the (angular) Compton wave-length, where $c$
is the velocity of light.  Then we would have obtained
$\E=mc^2/2$.)

However, for  classical orbits, it is often  convenient to
use {\it dynamical units}.  By these we mean units which depend on the
orbit itself and are determined by the initial conditions.
For our purposes it is convenient to choose the dynamical units
such that $a$ is the  turning
point and L is the corresponding angular momentum for the $E=0$ solution.
 The possibility of such a choice is based on
the following result:

\begin{quotation}
 {\it (a) Let  ``$a$" be a  prescribed, non-zero length scale. Then every
power-law potential, $V=-\gamma /r^\nu$  with $\gamma
>0$ and $\nu\ne2 $, will have an $E=0$ classical orbit  with
 $r=a$ as a turning radius. We call this orbit the ``standard orbit."

(b)This turning radius corresponds to the maximum (minimum) distance from
the center of the force for $\nu > 2$ ($\nu <2 $), respectively.}

(For $\nu=2$ the orbit has no turning point, except for the origin.)\\
\end{quotation}

Part (a) of the above result follows  from energy
conservation:
\be
 T+ V(r) =\left[ \frac{m}{2}\left(\frac{dr}{dt}\right)^2+\frac{L^2}{2mr^2}
\right] -\frac{\gamma}{r^\nu} =
\frac{m}{2}\left(\frac{dr}{dt}\right)^2+ U(L, r)=E~,
\lb{ect}
\ee
where  $L^2/2mr^2$ is the centripetal-barrier potential and
\be
 U(L,r) = \frac{L^2}{2mr^2}  - \frac{\gamma}{r^\nu}~ \lb{effpot1}
\ee
is the {\em effective potential}.
Since a turning radius corresponds to $dr/dt = 0$, the turning radius
$a$ must be a zero of $U(r)$ for $E=0$.
Thus, by combining Eqs. \rf{Dgamma} and \rf{effpot1}, we obtain
\be
U (L, r=a) = \frac{L^2}{2ma^2}-\frac{\gamma}{a^\nu}
   =\frac{L_0^2}{2ma^2}\left( \frac{L^2}{L_0^2} - g^2\right)=0~.
\label{turn}
\ee
Therefore, by choosing the angular momentum to be
\be
L=g L_0~,
\ee we
obtain the desired ``{\em standard orbit}."

By using  the dynamical unit $L$, instead of $L_0$,
 the potential $U$ is simplified to
\be
U(L,r)= \frac{L^2}{2ma^2}\left( \frac{1}{\r^2}-\frac{1}{\r^\nu}
\right)~, ~~~~~ \r=\frac{r}{a}~,        \lb{ur}
\ee
so that
\be
V(r)= -\frac{\gamma}{r^\nu}=- \frac{L^2}{2ma^2}\; \frac{1}{\r^\nu}=
- \frac{L^2a^{\nu-2}}{2m}\; \frac{1}{r^\nu}~.  \lb{Nvr}
\ee

Finally,  from
Eq.  \rf{Nvr}
we can read off the relation between $L$ and the
turning radius $a$:
\be
a= \left( \frac{2 m \gamma}{L^2} \right)^{\frac{1}{2\mu}}=
\left( \frac{2 m \gamma}{L^2} \right)^{\frac{1}{\nu-2}}~,~~~~~
\nu \neq 2~, ~~~\mu\ne 0~.     \lb{a}
\ee
In particular, for $L=0$ there is no turning point, except for
$a=\infty$. This special case is treated in Appendix A.

Part (b) of the result follows by substituting $E=0$ into Eq. \rf{ect}
 to yield
\be
 U(L,r)= -\frac{m}{2}\left(\frac{dr}{dt}\right)^2 \le 0 ~.
\ee
Combining this with Eq. \rf{ur} means that
\be
 \r^{\nu-2} \le 1~.
\ee
This condition determines the physical domain of the distance $r$:
\bea
  r &\leq& a~, ~~~~~ \nu > 2~,  \\  \nonumber
 r  &\geq& a~, ~~~~~ \nu < 2 ~.
\eea
Therefore, $a$ is  the apogee length
for bound orbits $(\nu > 2)$ and   the distance of closest
approach for unbound orbits $(\nu < 2)$.

For $\nu=2$ the above arguments break down, since the effective potential
$U=(L^2/2m-\gamma)r^{-2}$ does not allow a finite, non-zero turning point
when $E=0$.
This special case will be discussed further in Sec. 6.1.

Finally, we note that, having
established the existence of  a ``{\em standard orbit},"
it follows by  scaling arguments that there are an infinity
of {\em similar orbits}  whose turning radii
and angular momenta  are related to those
of the standard orbit as follows:
\be
  a_\lambda=\lambda a ~, ~~~~~~
L_\lambda=\lambda^{1-\nu/2} L =\lambda^{-\mu} L~.
\ee
These scaling relations are, of course, just special cases of the more
general $E\ne 0$ scaling law. (See Refs. \cite{landau,landau1} and
Appendix B.)

%***************************************************

\section{Classical Solutions }

\subsection{From the orbit equation}

A direct,  mathematically-inspired solution can be obtained by integrating
the orbit equation \cite{gold,gold1}
\be
\th - \th_0 = \int_{r_0}^{ r} \frac{r^{-1}\; dr}
        {\sqrt{\frac{2m}{L^2}(E-V)r^2 -1}}~.
\lb{Nthetaeq}
\ee
The general solution has the constant $\th_0$  in it.  We
presciently set $\th_0 = 0$ when $\rho = 1$ (the
turning-point condition), so that the turning point is along the positive
$x$-axis.

Then, setting $E=0$ makes the integral \rf{Nthetaeq}
doable for all $\nu$.  Changing variables successively
to $\rho=r/a$, $y=\r^{\mu}$, and $x=\cos^{-1}y$  means that
Eq. \rf{Nthetaeq} can be written as
\be
\th = \int_{1}^{\rho} \frac{\r^{(\mu-1)}~d\r}{\sqrt{1-\r^{2\mu}}}
= \mu^{-1} \int_{1}^{y} \frac{dy}{\sqrt{1-y^{2}}}
= -\mu^{-1} \int_{0}^{\cos^{-1}y} dx~.
\ee

Therefore, the solution is
\be
\r^{\mu}=y = \cos\left( - \mu\th \right)
      = \cos\left(\mu\th \right) ,
\ee
or
\be
\r = \left[\cos\left(\frac{(\nu -2)\th}{2}\right)
      \right]^{\frac{2}{\nu-2}}
    = \left[\cos\left(\mu\th\right)\right]^{1/\mu} . \lb{Nsolution}
\ee
We will discuss the allowed angular variations of $\th$ in the separate
sections on bound and unbound orbits.

\subsection{From the energy conservation equation}

We now give a  more intuitive derivation of the solution
\rf{Nsolution}.
By substituting the angular-momentum conservation condition
\be
\dot \th = L/(mr^2)    \lb{amc}
\ee
into the energy conservation condition
\be
E-V= \frac{m}{2} \dot \th^2
      \left[\left(\frac{dr}{d\th }\right)^2 +r^2\right] ~,
      \lb{ec}
\ee
one obtains \cite{problem,problem1}
\be
\left(\frac{dr}{d\th }\right)^2 +r^2 = \frac{2m(E-V) r^4 }{L^2}~.
\lb{Ncons}
\ee
This is essentially a first-order differential equation, which can
be formally integrated to yield the angular equation \rf{Nthetaeq}
of the last subsection.

However, for  $E=0$, it is much more
efficient to solve Eq.\rf{Ncons} directly.
Converting to the dimensionless variable $\r=r/a$ and substituting
$V$ from Eq. \rf{Nvr} into Eq. \rf{Ncons}, we obtain
 \be
\left(\frac{d\r}{d\th }\right)^2 +\r^2 = \r^{(4-\nu)}= \r^{(2-2\mu)} .
\lb{Nreq}
\ee
For $\nu = 4$ the right-hand side of this equation is unity, so the
solution is a cosine. This is the circular orbit $\r= \cos \th$
which we will discuss in detail in the next section.   Guided by this
and the substitution $y=\r^{\mu}$ of the last subsection,  we multiply
Eq. \rf{Nreq} by $\r^{2\mu-2}$ to yield
\be
\left(\r^{\mu-1}\frac{d\r}{d\th}\right)^2 +\r^{2\mu}=
\left(\frac{d\r^\mu}{\mu d\th}\right)^2 +\left(\r^\mu\right)^2 = 1 ~.
\lb{dift}
\ee
Now $\r^\mu$ satisfies the differential equation for
the trigonometric functions. Therefore, the {\em general} solution of Eq.
\rf{dift}
is  given by
\be
 \r^\mu = \cos \; \mu(\th-\th_0) = \cos\left[\frac{\nu -2}{2} (\th
-\th_0)\right]~,
\lb{sol}
\ee
or
\be
\r = \left[\cos\ \mu\left(\th-\th_0 \right)\right]^{1/\mu}=
 \left[\cos\left(\frac{\nu -2}{2} (\th-\th_0)\right)
      \right]^{\frac{2}{\nu-2}}~ .  \lb{solr}
\ee
The phase, $\th_0$, is the integration constant.
 It depends on
the initial conditions and, as before,
  we will set $\th_0$ to zero for our first-orbit turning-point condition.

%************************************************************

\section{Properties of the Bound Trajectories,
     $2<\nu$ or $1<\mu$}

 In Fig. 1 we show a typical effective potential $U(r)$ in this regime.
One sees that the value of $U(r)$ starts from $-\infty$ at
$r=0$, rises through zero
at $r=a$, reaches a maximum, and then decreases to zero at
$r \rightarrow \infty$.  Therefore, the
$E=0$ solution for $r=r(t)$ can only vary
between $r=0$ and $r=a$. We shall now discuss the dependence of $r$
on the azimuthal angle, $\th$.

\subsection{The first orbit}

The radius, $\r$, is nonnegative.
This means the range of the angle variable for the first orbit is
restricted.  Since, by convention, the first orbit has $\th_0 =
\th_0^1 = 0$,  this means
 $\th^1 = \th$ satisfies
\be
-\frac{\pi}{2\mu}= -\frac{\pi}{\nu-2}
\le \th^1 \le
\frac{\pi}{\nu-2}=\frac{\pi}{2\mu}~.
\ee

The corresponding
curve $\r=\r(\th^1)$ in the x-y plane,  begins at $\r=0$ for
$\th^1=-{\pi}/{2\mu}= -{\pi}/({\nu-2})$,  evolves counterclockwise to
$\r = 1$ at $\th^1 = 0$, and then continues on back to $\r = 0$ at
$\th^1 ={\pi}/(2\mu)=   {\pi}
/(\nu-2)$.  The curve evolves counterclockwise because we have chosen, by
convention, that the angular momentum is in the positive $z$ direction.
(Of course,
there is also a mathematically-rotated clockwise solution.)

Every such closed circuit, beginning and ending at the origin, we
will call an {\bf orbit}.  Because of the shapes of the orbits (which
we will present in the next Section),
when $\nu > 4$ we will also call an orbit a
{\bf petal} (of a flower)
and when $\nu < 4$ we will also call an orbit a
{\bf spiral}, which actually is a double spiral
since it spirals out and then
spirals in.  These two classes of orbits meet at $\nu = 4$, which
is the circle that goes through the origin.
 As we will see below, the entire
physical trajectory will consist of
 a pattern of  either (i) a finite integral number
of orbits which then repeat over themselves,
when   $\mu$ is a rational number,
or else (ii) an infinite number of nonrepeating orbits, if
$\mu$ is an irrational number.
(In Sec. 5 we will discuss the classically unbound orbits described by
 $\nu \leq 2$ or $\mu \leq 1$.)

\subsection{Solutions for later bound orbits}

\subsubsection{The angles and the phase shifts}

To obtain the entire physical solution, we must connect different orbits.
That is,
 the first orbit evolves into the second orbit...evolves into the $k$th
orbit.  Because the orbits go through the origin, both $\th$ and $\th_0$
are discontinuous, as we will demonstrate below.  Therefore, we will label
$\th$ and $\th_0$ for the $k$th orbit as  $\th^k$ and
$\th_0^k$, respectively.

The successive orbits must be connected
at the origin, $r=0$, in such a way that
the directions of the linear momentum $\hat{{\bf p}}$
and the angular momentum {\bf L} are continuous functions of time.
We {\em defined} the first orbit to have its apogee,  $r=a$, at $\th^1=0$,
which corresponded to the choice $\th_0^1=0$. In order to
obtain a physical solution, we must connect the first orbit to the second
orbit in such a way  that the tangent vector to the combined trajectory
is continuous at the origin.

Before the particle starts the second orbit, the trajectory goes though the
origin.
This creates a singular transition in the polar coordinates.
(Since the trajectory  passes through the origin and because
$\hat{{\bf p}}$ must be continuous, the direction $\hat{{\bf r}}$ of
the position vector must change sign at $r=0$.)
The angle advances by
$\pi$ (instead of $-\pi$) in going through the origin because, in order
to conserve angular momentum, the position vector must continue
rotating in the counterclockwise direction.
 Then the particle begins its second orbit,
travelling another angular distances $\pi/\mu$.  Before beginning its third
orbit, the particle goes through the origin again, advancing another $\pi$
radians.
Therefore, the angular variation of the $k$th orbit,
$\th^k$, is given  by
\be
\th^k_{min}\equiv\left(\frac{(k-3/2)}{\mu}+(k-1)\right)\pi
   \le \th^k
   \le  \left(\frac{(k-1/2)}{\mu}+(k-1)\right)\pi\equiv \th^k_{max}
\lb{Nthk}
\ee
and one has the condition
\be
\th^{k+1}_{min}=\th^{k}_{max}+ \pi ~.
\ee

However, the phase shift, $\th^k_0$, must also change with each orbit.
Recall, in summary, that
the $k$-th orbit is described in polar coordinates by
\be
\r_k = [\cos \mu (\th^k -\th^k_0)]^{1/\mu} ~,
\lb{dsol} \ee
with
\be
 |\th^k -\th^k_0|\le \pi/2\mu~
\lb{thk1}
\ee
or, equivalently,
\be
  \th^k_{min} \equiv \th^k_0-\frac{\pi}{2\mu} \le \th^k \le
\th^k_0+\frac{\pi}{2\mu}\equiv
\th^k_{max}~.
\lb{thk2}
\ee
Then,  from Eqs. \rf{Nthk} and \rf{thk2},
the phase shift, $\th_0^k$, is
\be
\th_0^k= \frac{ \th^k_{min}+\th^k_{max}}{2}
=(k-1)\left(\frac{\nu}{\nu-2}\right)\pi
= (k-1) \left(1+\frac{1}{\mu}\right)\pi~.
\ee

Mathematically this change of phase shift, $\th_0^k$, with each orbit
is because the orbit must be symmetric about
any apsidal vector \cite{gold2}.
This  is true even though the vector is
zero length from the origin.    (See the interesting
application of these concepts in the
$\nu=6$ orbit discussion of the next section.)
Further, the other
 apsidal vectors, ${\bf r}_k(\th^k=\th_0^k)$,
 are the symmetry axes of the $k$th orbits.

\subsubsection{Angle vs. phase-shift differences}

It is instructive to observe that, instead of using the label, $k$, for
each orbit, formally we can
write the general solution for all times $t\ge 0$ in the form
\be
\r(t) = [\; \cos \mu ( \th(t)-\th_0(t))\; ]^{1/\mu}\equiv
[\; \cos \mu \chi(t)\; ]^{1/\mu} ~, \lb{form}
\ee
with
\be
 \chi (t) \equiv \th(t)-\th_0(t)~.
\ee
Here, both the azimuthal angle, $\th$, and
also the phase shift, $\th_0$,
are  regarded as
functions of time.  These angles have discrete jumps at the crossing times
\be
t_k\equiv (k-1/2)\tau~,  \lb{crt}
\ee
where  the $t_k$ are the times
at which the particle passes through the origin
and $\tau$ is the period of one orbit.
Using our same physical convention, which here is $r(0)=a$ at $t=0$ with
$\th(0)=\th_0(0)=0$, we have that the jumps are
\be
\th(t_k +\epsilon) =\th(t_k -\epsilon)+\pi~
\ee
and
\be
\th_0(t)=\pi  \left(1+\frac{1}{\mu}\right) \sum_{k=1}^\infty
\Theta(t-t_k)~, \lb{tkt}
\ee
where $\Theta(t)$ is the Heaviside step function.

We see that both $\th$ and $\th_0$ are  monotonically increasing functions
of time.   In contrast, their
difference, $\chi(t)$, is a  periodic function of $t$. For every new orbit,
$\chi$ starts with the value $-\pi/(2\mu)$  and  then increases
monotonically to $\pi/(2\mu)$,
 due to the continuous increase in $\th(t)$
as the particle goes through one orbit (petal or spiral).
Next, $\chi$
decreases stepwise by $\pi/\mu$ at the crossing time. Therefore, as
in our previous representation,
$\cos \mu \chi$ starts and ends each orbit at $\r = 0$.

\subsection{Precession of the orbits}

Because of  angular-momentum conservation,
the azimuthal angle, $\th$, is a monotonic function of time, $t$.
If there is no precession,  $\th$ {\em increases} exactly
by $2\pi$ after one period, $\tau$. However, if after one period
the axis of the
orbit has rotated {\em forwards or backwards}
(i.e.  clockwise or counterclockwise) from $2\pi$ by
an angle $P_\nu$,  then this is the precession per period.
Specifically,
\be
 \th(t+\tau)-\th(t)= 2\pi + P_\nu ~, \lb{defp}
\ee
where $\tau$ is the period of one orbit. If we choose $t$ to be
the time at which the particle was at the apogee of the $k$th orbit,
then $t+\tau$ will be the time at which the particle reaches the
apogee of the $(k+1)$th orbit. To do this, the particle must
first rotate by $\pi/2\mu$ to reach the origin, have
its angle $\th$ jump by
$\pi$ at the
origin, and finally rotate by another $\pi/2\mu$ to reach the new apogee.
Thus, from Eq.\rf{defp} we have
\be
\th(t+\tau)-\th(t)= \frac{\pi}{2\mu} + \pi + \frac{\pi}{2\mu}=
 \left(\frac{1}{\mu} + 1\right) \pi =
2 \pi + P_\nu~,
\ee
so that
\be
P_{\nu}=  \left(\frac{1}{\mu}-1\right)\pi=
\left(\frac{4-\nu}{\nu-2}\right)\pi~.   \lb{P}
\ee
In Fig. 2 we plot $P_{\nu}$ as a function of $\nu$.  The
precession  is infinite at $\nu = 2$,  decreases through  zero at
$\nu=4$, and asymptotes to $-\pi$ as $\nu$ goes to infinity.
Therefore,  the
precession will be zero for $\nu=4$, negative (counter-clockwise)  for
$\nu>4$, and positive (clockwise) for $\nu<4$.

This also demonstrates that, for $\nu$ equal to a rational fraction, the
orbits will eventually close on themselves, and repeat.  This happens when
$kP_{\nu}$ is an integer times $2\pi$.

\subsection{Classical period}

The classical period, $\tau$, can be obtained by integrating the
angular-velocity equation
\be
\frac{d\th}{dt} = \frac{L}{mr^2} = \frac{1}{\tau_0 \r^2}~,
\label{ave}
\ee
where $\tau_0$ is a convenient unit of time,
\be
   \tau_0 = \frac{ma^2}{L}~.
\ee
One then has that the classical period in units of $\tau_0$ is
\be
T_{\nu} = \frac{\tau_{\nu}}{\tau_0}
=\int_{\frac{-\pi}{|\nu-2|}}^{\frac{\pi}{|\nu-2|}} \r^2 (\th)~d\th
= \int_{\frac{-\pi}{2|\mu|}}^{\frac{\pi}{2|\mu|}}
         \left[\cos(|\mu| \th)\right]
        ^{\left(\frac{2}{\mu}\right)}~d\th~.  \label{abs}
\ee
(In Eq. (\ref{abs}) we have inserted absolute values around $\mu$ in the
arguments of the integral and integrand.  This allows us to include
the negative $\mu$ case that we will return to at the end of this
discussion.)

Changing variables to $x = |\mu| \th$ allows
the integral to be evaluated as \cite{gr}
\bea
T_{\nu} & = & \frac{2}{|\mu|}\int_{0}^{\frac{\pi}{2}}
         \left[\cos x\right]^{\left(\frac{2}{\mu}\right)}dx
   =  \frac{1}{|\mu|} B(1/2,~b)   \\
   &=& \frac{\sqrt{\pi}}{|\mu|}
        \frac{\Gamma(b)}{\Gamma(b + 1/2)} ~, ~~~~~ b>0~,
          \lb{Ngama}
\eea
where
\be
 b \equiv \frac{1}{\mu}+\frac{1}{2} = \frac{\nu + 2}{2\nu - 4}>0~.
\label{b}
\ee
 $B(b,c)=\Gamma(b)\Gamma(c)/\Gamma(b+c)$ is the beta function and we
used $\Gamma(1/2)=\sqrt{\pi}$.
We see that $T_\nu$ is finite and well-defined
for $\mu>0$ or $\nu>2$, as would be expected for  bound
orbits.

In Fig. 3 we plot $T_{\nu}$
as a decreasing  function of $\nu$.  From
Eq. \rf{Ngama}  special cases are
\bea
\lim_{\epsilon \rightarrow 0} T_{2+\epsilon} &=&
      \left(\frac{2\pi}{\epsilon}\right)^{1/2} ~,  \label{case} \\
T_3 &=& \frac{3\pi}{4} ~,  \nonumber \\
T_4 &=& \frac{\pi}{2}~,  \nonumber  \\
T_6 &=& 1~,  \nonumber   \\
\lim_{\nu \rightarrow \infty} T_{\nu} &=&  \frac{2\pi}{\nu}
{}~.     \nonumber
\eea
The first equality in Eq. (\ref{case})  is obtained by using
\cite{bate} the
relation $\Gamma(z+1/2)\simeq \Gamma(z) \sqrt{z}$, which holds for $|z|
\rightarrow \infty$.

In passing, note that Eqs. \rf{Ngama} and (\ref{b})
also tell us that $T_\nu$ is  finite and well-defined for
the unbound, infinite orbits corresponding to $\mu <-2 $
or $\nu <-2$. This means that
it takes a finite time
for the particle to travel in from infinity, reach
the turning point $r=a$, and then go back to infinity.

\subsection{Geometric approximation to the area of the orbit}

Kepler's 2nd Law states that equal areas are swept out in equal times.
Although usually considered for the gravitational potential,
it actually holds for any central potential.
The proportionality constant follows from the
angular-momentum conservation  equation \rf{amc}:
\be
d{\cal A} = \frac{1}{2} r^2 d\th
  =\frac{1}{2} r^2\dot\th\; dt=\frac{L}{2m} \; dt~,
\lb{knd}
\ee
which is related to the angular-velocity equation (\ref{ave}) in an
obvious way.
Integrating this equation over a complete period, we get
\be
{\cal A}_{\nu}=\frac{L}{2m}\tau_{\nu}~,
\ee
where once again the subscript $\nu$ labels the potential we are considering.
 We now introduce a dimensionless area, $A_{\nu}$, as
the area of an orbit in units of $a^2$:
\be
A_{\nu} \equiv   \frac{{\cal A}_{\nu}}{a^2}= \frac{L}{2ma^2}  \tau_{\nu} =
 \frac{\tau_{\nu}}{2\tau_0}  \equiv \frac{1}{2} T_{\nu} ~.
\ee
Thus,  the reduced  area, $ A_{\nu}$, is
  equal to one-half  the reduced period, $T_{\nu}$.

For these $\nu > 2$ or $\mu > 0$ cases,
we can give amusing,
geometric, upper and lower bounds to the areas of the orbits.   Recalling
that the opening angle of a petal or spiral is
\be
\Phi_\nu =  \frac{\pi}{\mu}= \frac{2\pi}{\nu - 2}~,
\ee
an upper bound is defined by the area of section of a unit circle with opening
angle $\Phi_\nu$.  This bounds the real area as
\be
A_{\nu} < \frac{1}{2}~\Phi_{\nu}~.
\ee
Of course, this sectional area is equal to half the opening angle itself,
measured in radians.

We can also obtain another bound, by considering the quantity
\be
\tilde \r(\th) \equiv \cos \mu \th
\ee
in comparison to the real orbital radius
\be
 \r = [\cos \mu \th]^{1/\mu}~.
\ee
By using $\tilde \r(\th)$ in place of $\r$ in the integral for $A_{\nu}$,
\be
 A_{\nu} =
\frac{1}{2}\int_{\frac{-\pi}{2\mu}}^{\frac{\pi}{2\mu}} \r^2 (\th)~d\th
=\frac{1}{2} \int_{\frac{-\pi}{2\mu}}^{\frac{\pi}{2\mu}}
         \left[\cos \mu \th \right]^{\frac{2}{\mu}}~d\th~,
\ee
one obtains the quantity
\be
 {\tilde{A}}_{\nu} \equiv
\frac{1}{2}\int_{\frac{-\pi}{2\mu}}^{\frac{\pi}{2\mu}} \tilde \r^2 (\th)~d\th
= \Phi_{\nu}/4~.
\ee
${\tilde{A}}_{\nu}$ is an upper bound of the
reduced area when $2 < \nu \le 4$ or
$0 < \mu \le 1$, because then $\tilde \r(\th)$ will be outside the area of
the orbit.  However, for $4 \le \nu $ or
$1 \le \mu $,
$\tilde{A}_\nu $ becomes a lower bound, because then $\tilde \r(\th)$
is inside the  orbit.

\indent From the above, we can make special case
comparisons of these bounds, similar to the results of Eq.(\ref{case}).
Differing by a factor of $2$ because of the reduced area definition,
and written in the form ``lower bound" $\le$ ``exact reduced area"
$\le$ ``upper bound," one has
\bea
\lim_{\epsilon \rightarrow 0}
 {A}_{2+\epsilon}&=&
      \left(\frac{\pi}{2 \epsilon}\right)^{1/2} <
      \frac{\pi}{2\epsilon} = \frac{1}{4} \Phi_{2+\epsilon}~,  \lb{case2} \\
{A}_{3}&=&\frac{3\pi}{8} <
\frac{\pi}{2} =\frac{1}{4} \Phi_{3} ~,   \nonumber \\
\frac{1}{4}\Phi_4=\frac{\pi}{4} ={A}_{4}&=&
\frac{\pi}{4} = \frac{1}{4}\Phi_4
<  \frac{\pi}{2} =\frac{1}{2} \Phi_{4}~,  \nonumber  \\
\frac{1}{4} \Phi_{6}= \frac{\pi}{8}<{A}_{6}&=&
\frac{1}{2}<\frac{\pi}{4} =
\frac{1}{2}\Phi_{6}~,   \nonumber  \\
\frac{1}{4} \Phi_{\nu} = \frac{\pi}{2(\nu-2)} <
\lim_{\nu \rightarrow \infty}{A}_{\nu}
&=& \frac{\pi}{\nu}
< \frac{\pi}{\nu-2}= \frac{1}{2} \Phi_{\nu}~.  \nonumber
\eea

The upper bound
$\frac{1}{2}\Phi_{\nu}$ is very good as $\nu \rightarrow \infty$,
because the area of the orbit is a very thin petal.  As $\nu$ becomes
smaller, the approximation becomes worse.  By the time  $\nu=4$
or $\mu = 1$ this upper bound is off by a factor of $2$.
This is because the reduced area is that of
a circle of radius $1/2$, whereas this upper bound is equal to half
the area of a unit circle.  Both
upper bounds become useless as $\nu \rightarrow 2$.  This is because  the
upper bounds $\Phi_\nu/4$ and $\Phi_\nu/2$ are describing  areas
with  angular widths approaching infinity and radius $1$, whereas the true
area has a radius that is spiraling in to zero as the angular position
becomes large.

Finally, note that the lower bound goes from being off by a factor of $2$ as
$\nu \rightarrow \infty$ to equality at $\nu = 4$.

%**********************************************

\section{Examples of  Classical Bound Trajectories}

\subsection{Petals: $\nu > 4$ or $\mu > 1$}

\noindent From the previous section we see that, as $\nu$ approaches
infinity, the orbit is an ever-narrowing petal of width
\be
\Phi_\nu =  \frac{\pi}{\mu}= \frac{2\pi}{\nu - 2}~.
\ee
Similarly, the counterclockwise precession
reaches $-\pi$ per orbit as $\nu \rightarrow \infty$.  As $\nu$ becomes
smaller, the petals increase in width.  We illustrate this with
 several illuminating examples.

$\nu=8$, {\em three petals:}\ We begin with the case $\nu=8$.
Here a petal is  $\pi/3$ wide
and the precession per orbit is $-2\pi/3$.  Thus, there are three orbits
before the trajectory closes. Note that the three petals in a closed
trajectory  cover only half of the opening angle from the origin.
We show this in Figure 4.

$\nu=7$, {\em ten petals:}\ For $\nu=7$ a petal is $2\pi/5$ wide
and the precession per orbit is
$-3\pi/5$.  This means that, before the trajectory closes,
 there must be ten orbits  and
the precession goes around three times.

$\nu=6$, {\em perpendicular lemniscates:}\
The case $\nu=6$ is very interesting.
 The width of a petal is $\pi/2$ and
the precession is $-\pi/2$ per orbit. Here, the width of a petal and the
precession are exactly such that there is
no overlap and also no ``empty angles."
It  takes four orbits to
close a trajectory.  This is shown in Figure 5. We see that the physical
solution consists of two perpendicular lemniscates (figure-eight
curves composed of two opposite petals).
This conclusion is in contradiction
to one given in the literature \cite{problem3}. There, a single
lemniscate was predicted.  From the physical arguments we have given
above, this is not valid.
This incorrect solution resulted from studying the equation
for the square of the orbit.

$\nu = 5$, {\em six overlapping petals:} The petals are $2\pi/3$
wide and they
precess by $-\pi/3$ per orbit. This means there are
six petals in a closed trajectory (flower).  Note that here a petal
is still so wide and the precession is so small, that the successive
petals overlap.

\subsection{Circle through the origin: $\nu = 4$ or $\mu = 1$}

For $\nu=4$, the solution is well known  \cite{problem2}.  It is a circle
that starts at the origin, travels symmetrically about the positive $x$-axis,
and returns to the origin.  The precession is zero,  so the orbit
continually repeats itself.  In Fig. 6 we show the orbit.

\subsection{Spirals:\ $2 <\nu < 4$ or $0< \mu < 1$}

As $\nu$ becomes less than $4$, we can think of a petal obtaining a width
greater than $\pi$, i.e.,  an orbit consists of
two spirals, one out and one in,  at opposite ends of
the orbit.   As $\nu$ approaches $2$.
the spirals become tighter and tighter and the precession (now clockwise)
becomes larger.
In fact, the spirals' angular variation as well as the orbit's precession
both become infinite in magnitude as $\nu$ approaches $2$.

Consider the special case $\nu=3$.  The width of the double-spiral  orbit
is still given by the formula for
$\Phi_{\nu}$, and is $2\pi$.  Therefore, the first orbit
begins and ends towards the negative $x$-axis.  The precession is $\pi$,
so the trajectory closes after two orbits.  We show this case in Figure 7.

When $\nu=7/3$, the width of the double-spiral is $6\pi$.
This means the first orbit starts towards the negative $x$-axis,  winds
around one and a half times before reaching the positive axis, and then
winds one and a half times more to reach the origin again.  The precession
is $5\pi$, so the trajectory closes after two orbits.  We show this case
in Figure 8.  However, on this scale it is impossible to see the tight
winding of the spirals near the origin.  Therefore, we
demonstrate it  with expanded views in Fig. 9.

As $\nu \rightarrow 2$ from above, the spirals get tighter and
tighter.  Their angular widths, $\Phi_\nu=2\pi/|\nu-2|$, become larger and
larger as $\nu \rightarrow 2$.

%************************************************************

\section{Unbound Classical Trajectories,  $\nu \le 2$ or $\mu \leq 0$}

\subsection{The infinite spiral:\ $\nu=2$ or $ \mu=0$}

When $\nu$ reaches $2$, there is a singular change.  First, the
double spiral becomes infinite in angular width. But also, the joining of
the two sides of the
double spiral at $\r=1$ and $\th=0$ breaks down.   It is as if a
tightly-wound double spring broke.  The ends
spiral out to infinity, as we discuss below.

The $\nu=2$ is a special case in other ways.
For us, with our $E=0$ condition,
it is the demarcation between the bound orbits, $\nu > 2$, and the unbound
orbits, $\nu < 2$.  But also, for $\nu \equiv 2$ and $E \neq 0$, it is the
demarcation between the $E<0$ bound solutions and $E>0$ unbound solutions.
These latter are Cotes' spirals \cite{marion}.  The $E\ne 0$ solutions are
given in Ref. \cite{cotes}.

Returning to $E=0$, Eq. \rf{effpot1} shows that,
for $\mu=0$ or $\nu=2$, there is no natural length scale.
This is because both the centripetal and the external potentials
have the same power dependence on $r$:
\be
U(r)=(L^2/2m-\gamma)\frac{1}{r^2}~.
\ee
Thus,  for all $r$, the effective potential $U(r)$ is either repulsive
(and positive-valued) or else it is
attractive (and negative-valued).

For $U(r)>0$
there is no solution for $E=0$. However, for $U <0$ the
energy-conservation condition \rf{ec} with $E=0$ gives
\be
 \left(\frac{dr}{d\th}\right)^2=\left(2m \gamma/L^2-1\right) r^2\equiv
\lambda^2
 r^2 ~, \lb{drdf}
\ee
so that
\be
{dr}/{d\th}=\mp \lambda r~.
\ee
Therefore, the two possible
solutions are
\be
r = r_0 ~\exp\; [\mp \lambda \; (\th-\th_0)]~,
{}~~~~~ -\infty<\th< \infty~.
\lb{r}
\ee
These correspond to orbits which pass through the point $(r_0,\th_0)$
and spiral inwards or outwards for the
 minus or plus signs, respectively.
For example, a complete, counterclockwise
orbit, starting from the initial value $r=r_{in}=r(\th_{in})$, going
into  the origin,  and
then going out of the origin
towards the final value $r=r_{f}=r(\th_f)$,
is given by
\bea
\frac{r}{r_{in}} &=& \exp\; [- \lambda \; (\th-\th_{in})] ~,
                 \qquad  \mbox{ $\th_{in} \leq \th < \infty$}~,
\lb{rspirali}
\\
 \frac{r}{r_{f}} &=& \exp\; [+ \lambda \; (\th-\th_f)] ~,
                 \qquad  \mbox{ $-\infty < \th \leq \th_{f}$}~.
\lb{rspiral}
\eea
However, it is important to note that Eqs. \rf{rspirali} and \rf{rspiral}
are only special cases of Eq. \rf{r}. For example,
Eq. \rf{r}, with the plus sign,
can also describe an  $r$ starting at $r(\th_0)=r_0$ which then
goes out to infinity as $\th \rightarrow \infty$.

Despite the infinite spirals in the  example of Eqs. \rf{rspirali} and
\rf{rspiral},
the journey in and out takes only a
finite time.  To see this, consider
the time dependence of $r$. From the energy-conservation
conditions \rf{ect} and \rf{drdf},
we have
\be
\dot r^2= \frac{2}{m} \frac{\gamma}{r^2}
- \frac{L^2}{m^2 r^2} =\frac{L^2}{m^2}\left(\frac{2m \gamma}{L^2}-
 1 \right)  \frac{1}{r^2}= \left(\frac{L\lambda}{m}\right)^2 \frac{1}{r^2}~.
\ee
Therefore,
\be
\frac{d r^2}{dt}= 2 r \dot r  =  \mp \frac{2L\lambda}{m}~
\ee
and
\be
r^2=r_0^2 \mp \frac{2L\lambda}{m} (t-t_0)~.  \lb{rt}
\ee
This tells us that, in spite of  the infinite spiralling, a
particle moving {\em inwards} reaches the origin, $r=0$, from any point
$r_0$ in a finite time, given by
\be
 \Delta t =t-t_0 =  \frac{m}{2L\lambda} \; r_0^2~.
\ee
Finally, substituting Eq. \rf{r} into Eq. \rf{rt} gives the
general time dependence of
$\th$ as
\be
t-t_0 =\mp \frac{m}{2L\lambda} (r^2- r_0^2 )
=\mp \frac{m r_0^2 }{2L\lambda}\{\exp [\mp 2\lambda
(\th-\th_0)] -1 \} ~.  \label{lll}
\ee
Because of our choice ${\bf \hat L}={\bf \hat z}$,
in Eq. (\ref{lll}) we are assuming that $\th$ increases monotonically with
time.

\subsection{General unbound orbits:  $\nu < 2$ or $\mu < 0$}

When the potential
 parameter $\nu$ just leaves
that of the infinite spiral, that is, when
one barely has $\nu < 2$ or $\mu < 0$, there is another change.  Although
the two ends of the entire trajectory still reach to infinity and the
spirals in and out almost have  infinite angular widths, the distance of
closest approach jumps from $\r = 0$ to $\r = 1$.

As  the value of $\nu$ decreases, the value of the angular width
of the trajectory, now
given by  $\Phi_{\nu} = \pi/|\mu|$, also decreases accordingly.
By the time $\nu = 1$,
the angular width has decreased to $2\pi$.    Eventually it
becomes less than $\pi$, meaning the orbit comes in and
out in the same half plane.  That is, when $\nu < 0$, the force is
repulsive.  In the next section we give some illustrative examples.

%**************************************************

\section{Examples of Classical Unbound Trajectories}

\subsection{Kepler-like potentials:\ $0 <\nu < 2$ or $-1 < \mu < 0$}

When  $0 < \nu < 2$, the repulsive
centripetal barrier dominates at small
$r$ whereas the attractive potential  $V=-\gamma/r^\nu$
dominates at large
$r$. A typical shape of the effective potential is sketched
in Fig. 10.  It is familiar from the Kepler problem.
Therefore, for $0 < \nu < 2$, the $E=0$ classical orbits are all unbounded.
The distance, $a$, defined in Eq. \rf{a} now has a completely
different interpretation.  It is now  the
distance of closest approach.  Even so, the formal solution \rf{sol} remains
valid for negative values of $\mu$.

As a first example consider the case $\nu = 3/2$ or $\mu = -1/4$.
This orbit has a total angular width of $4\pi$.  It is shown in the
two drawings of Fig 11.  The large-scale first drawing shows the
trajectory coming in from the top, performing some gyration, and going
out at the bottom.  The small-scale second drawing shows the trajectory
winding around twice near the origin, with the distance of closest
approach being one.

A second example is the  exact Kepler potential,
$\nu=1$ or $\mu=-1/2$.
Eq.  \rf{sol} gives
\be
\r^{-1/2}=\cos \th /2 ~,
\ee
so that
\be
\frac{1}{\r}=(\cos \th /2)^2=\frac{1+ \cos \th}{2} ~.
\ee
This is the famous  parabolic orbit for the Kepler
problem with $E=0$.   This orbit is shown in the first drawing
of Figure 12.  The parabola yields an angular width of $2\pi$, as it
should.

\subsection{The straight line: $\nu = 0$ or $ \mu = -1 $}

If we formally set $\nu=0$ in the expression \rf{potN}, we get a
negative constant potential $V(r)=-\gamma$. Therefore,
in this case the force vanishes
 and we have a free particle.  Its  orbit must be a
straight line.  However, Eq. \rf{Nreq} shows that one still has
the same type of  solution, Eq. \rf{solr}.  Here
it is
\be
\r = [\cos\th]^{-1}~, ~~~~~ x = r\cos\th=a ~.
\ee
This is the equation for a vertical straight line that crosses the
$x$-axis at  $x=a$, as required by the initial conditions.
This orbit is shown in the second drawing Fig. 12, it subtending an
angular width of $\pi$ from the origin.

\subsection{Repulsive potentials:
$-\infty < \nu < 0$ or $-\infty < \mu < -1 $}

For $\nu <0$ the potentials $V(r)$ in Eq. \rf{potN} are repulsive and
negative-valued for all $r>0$,  with $V(r)$ going to $-\infty$ at large
distances.   Since both the potential, $V(r)$, and the centripetal potential
decrease monotonically, the effective potential has no minima or maxima.
Even so,  for $E=0$ these unbounded orbits  behave qualitatively like those
for $ 0 \leq \nu < 2$. The distance of closest approach again obeys the
formula \rf{a} and the solutions are given by the same expression
\rf{sol}, which is valid for all $\mu \ne 0$.

The  solution \rf{solr} for the first orbit ($\th_0^1=0$) gives
\be
\r=[\cos\mu \th]^{1/\mu}= [\cos |\mu| \th]^{-1/|\mu|}~.
\ee
This shows that
  the particle is at $r=a$ when $\th=0$ and it is at  infinity
when $ \th=\pm\pi/(2|\mu|) $. Thus, these orbits become narrower as
$|\mu|$ becomes larger.  This is similar to the case of the width of
the petals of the bound orbits.

{\em Hyperbolic orbits:}\ The most famous  special case of these
potentials is the ``inverted"
harmonic-oscillator potential, with $\nu=\mu=-2$.
The orbit  is given by $\r= [\cos 2\th]^{-1/2}~$, so that
\be
1 = \frac {r^2}{a^2} \cos 2\th=\frac {r^2}{a^2}
(\cos^2 \th - \sin^2 \th )= \frac {x^2}{a^2}- \frac {y^2}{a^2} ~.
\ee
Thus, the trajectory
is a  special hyperbolic orbit, whose minor
and major axes are equal, $b^2=a^2$. (In fact,
every $\nu=-2$  solution, i.e. for arbitrary  $E\ne 0$, also
yields a  hyperbolic orbit, but with $b \ne a$.)
We show this orbit as the third drawing in Fig. 12.  Now the
angular width has decreased to $\pi/2$.

As the last case, we
consider  the orbit for $\nu = -4$ or $\mu = -3$.  This orbit is shown
in the last drawing of Fig. 12.  The orbit subtends an angle of
$\pi/3$, again as it should.  One sees that as $\nu$ becomes more and
more negative, the orbits will become narrower and narrower.  This is
just as in the bound case, where the petals became narrower and
narrower as $\nu$ became more and more positive.

%*****************************************************

\section{Comments on the Classical Problem}

It is well-known that classical orbits precess  for general central
potentials. There are two  exceptions, the
Kepler and the isotropic harmonic oscillator potentials \cite{bertrand}.
Further, these central potentials are known  to have additional conserved
quantities:  the Runge-Lenz vector, ${\bf A}$, in the Kepler problem and
the quadrupole moment tensor, $Q_{ij}$, in
the harmonic oscillator problem
\cite{moregold,schiff}.
These two examples show that there is a close connection between the
absence of precession and the existence of dynamically conserved
quantities.

On the other hand, it is also well-known that for a specific energy to
angular-momentum combination, general central potentials can have
closed orbits.  But these combinations have different unique values,
depending upon the potential.

Therefore, it is interesting that the
class of $E=0$ solutions for the potentials \rf{pot} with $\nu > 2$, have a
countable number of examples with well-defined, closed trajectories.  They
are  all the potentials with rational values
of $\nu$. One of them, the case $\nu = 4$, has zero precession.

All the $E=0$ with  $\nu > 2$ solutions are bound and
 go through the origin.
The $\nu < 2$ solutions are unbound, and go to
infinity.  (The $\nu = 2$
 case is special, and does both.)

We have seen that, with the $\nu > 2$ singular potentials, the physical
solutions  differ  from more familiar solutions in that,
at the origin, we must   paste together single-orbit
 solutions  which have different
phase-shifts, $\th^k_0$, at $r=0$.
The   discontinuity at the origin of the phase shifts,
given by Eq. \rf{tkt}, is
fundamental \cite{disc}.
It is necessary because the bound orbits pass through the origin, where
the potential is singular.  This causes
the second order time-derivative of the position, i.e., the acceleration,
to be infinite at $r=0$ when  $\nu \ge 2$:
\be
\ddot x_i(t) = - \frac{\nu\gamma}{mr^{\nu+2}} x_i ~, \qquad i=1,2,3 ~.
\ee
Therefore,  irrespective of what coordinate system we choose,
at the origin we must
still paste together  different solutions of Newton's
equation which depend on different initial conditions, as exhibited in
Eq. \rf{dsol}.

As we have just observed, when crossing the center of singular
potentials, a particle will have infinite momentum and kinetic energy.
This leads to a violation of the virial
theorem, even for these bound trajectories.
Formally applying  the virial theorem \cite{goldV,landauV} to power
potentials leads to the following well-known equation:
\be
\sp{T} = - \frac{\nu}{2} \sp{V}~,    \lb{vir}
\ee
where $\sp{T}$ and $\sp{V}$ are the time averages of the kinetic
and potential energies. The above equation leads
to two famous relations:
$\sp{T} = - \sp{V}/2 $, for  Kepler  elliptical orbits
($E<0$  and $\nu=1$), and $\sp{T} =\sp{V}$ for the harmonic
oscillator ($E>0$ and $\nu=-2$).

\indent From Eq. \rf{vir} and energy conservation, we have
\be
  E=T+V= \sp{T} +\sp{V}=\left(1-\frac{2}{\nu}\right) \sp{T}~.  \lb{vir2}
\ee
Since $ \sp{T}\;  > 0$, Eq. \rf{vir2} immediately shows that  the
virial theorem \rf{vir} is violated by all the $E=0$ solutions of power
potentials with $\nu \ne 2$.  Such a violation of \rf{vir} is
expected for infinite orbits, such as the parabolic
orbit of the Kepler problem and all other $E = 0,~ \nu < 2$ orbits.
What is of more interest is that the bound solutions,
$\nu > 2$, also violate the
theorem.  In Appendix C we will go into more detail on this point

In summary, the classical $E=0$ solutions for the power-law potentials
exhibit a
fascinating set of properties.  In paper II we will show that this
characteristic is also true for the  quantum solutions.

\newpage

%**************************************************

{\large{\bf  Appendix A: Zero Angular-Momentum Solutions}} \\

Here we discuss the simpler $E=L=0$ solutions, where
the effective potential $U(r)$ is equal to the potential $V(r)$ itself.
A particle  moves radially, in a straight line,  between zero and
infinity, with no turning points at
 any finite radius, $r\ne 0$.

\indent From the energy-conservation
equation \rf{ect} one has that, for $E=L=0$,
\be
\frac{dr}{dt} =~\pm~ \left(\frac{2\gamma}{m}\right)^{1/2}
                \frac{1}{r^{\nu/2}} ~
                \equiv ~ v(t)~. \lb{econ}
\ee
Unless $\nu = -2$ this is equivalent to
\be
\frac{d(r)^{\nu/2+1}}{(\nu/2+1)dt} =
         ~\pm~ \left(\frac{2\gamma}{m}\right)^{1/2}~,
\ee
which has the solutions
\be
r(t) = \left[r_0^{\frac{\nu+2}{2}}\pm~ \left(\frac{2\gamma}{m}\right)^{1/2}
      \left(\frac{\nu}{2}+1\right)(t-t_0)\right]^{\frac{2}{\nu+2}},
           ~~~~~~\nu \neq -2.  \lb{z}
\ee
where $r_0\equiv r(t_0)$.

The solution \rf{z} does not depend explicitly on
the initial velocity $v_0 \equiv v(t_0)$ because $r_0$ and
$v_0$ are not independent.
However, because of the condition $E=0$,   using Eq. \rf{econ}
yields  $\pm~ (2\gamma/m)^{1/2}=v_0 r_0^{\nu/2}$.
Substituting this  relation into Eq. \rf{z} and expanding yields
\be
r(t) = r_0 + v_0 (t-t_0) - \frac{\nu}{4} \frac{v_0^2}{r_0} (t-t_0)^2
+ \frac{\nu(\nu+1)}{12} \frac{v_0^3}{r_0^2} (t-t_0)^3 +
\cdots~.\lb{expr}
 \ee
This expansion is absolutely convergent for $|v_0 (t-t_0)| < r_0$.
This power series terminates, if the index $\nu$ is equal to negative
integers, $\nu=0,-1,-2,\cdots$. In particular, for a constant force,
$\nu=-1$, we get the familiar result,
$r(t)= r_0 + v_0 (t-t_0) + \frac{1}{2}a (t-t_0)^2$, where the constant
acceleration is given by $ a=\gamma/m= v_0^2/(2 r_0)$.

Eq. \rf{z} shows that for $\nu > -2$ it takes
 an infinite time to travel between $r_0\ne 0$ and $r=\infty$, but
only a finite amount of time to reach the origin $r=0$. This
finite arrival time is given by
\be \Delta t \equiv t-t_0= \frac{r_0}{|(\nu/2+1) \ v_0|}~. \lb{artime}\ee
Note that $\Delta t$ is equal to the time
required by a particle moving with {\em an effective constant velocity},
\be    v_{eff} = |(\nu/2+1) \ v_0|~, \lb{veff}\ee
to reach the origin from $r=r_0$.
In particular, for the free particle, $\nu = 0$, the above effective speed
\rf{veff} is equal to the actual constant speed $v_0$, as it should.

However, for $\nu < -2$, the power of the
large square brackets in Eq. \rf{z} is negative.
This yields the opposite situation: Now, it  takes an infinite time to
reach $r=0$ from any $r_0\neq 0$. But it takes a finite time, given by
\rf{artime}, to travel between any $r_0 \neq 0$ and infinity. The latter
result is consistent with the result of Sec. 4.4, that
 this travel time is  finite  for $L\ne 0$.

Finally, the special case $\nu = -2$ yields the differential equation
\be
\frac{dr}{dt} =~\pm~ \left(\frac{2\gamma}{m}\right)^{1/2}r~.
\ee
The solutions are therefore,
\be
r(t) = r_0 ~\exp\left[\pm~ \left(\frac{2\gamma}{m}\right)^{1/2}(t-t_0)\right]
= r_0~ \exp\left[ \left(\frac{v_0}{r_0}\right)(t-t_0)\right]~. \lb{exps}
\ee
The Taylor expansion agrees with Eq. \rf{expr} with $\nu=-2$. In fact, the
exponential solution \rf{exps} follows from the general solution \rf{z},
by  using
 $ \lim_{\epsilon \rai}(1+\epsilon x)^{1/\epsilon}= \exp [x] $.
Note that for $\nu=-2$ the travel time
between  $r_0\ne 0$ and either
$r=0$ or $r=\infty$ becomes infinite. This is consistent with  $\nu=-2$
being a transition between  the regimes where the travel time is
infinite to reach
the origin ($\nu < -2$) or infinity ($\nu > -2$).

Therefore, for the $L=0$ case, the complete journey from the origin
to infinity takes an infinite amount of time for {\it all}
indices $-\infty < \nu < \infty$. \\

%**************************************************

{\large{\bf  Appendix B: Similar Orbits and Scaling}} \\

The power potentials have the property
 that for each solution, $r_1= a \r(\th)$,
there are infinitely many {\em similar solutions}, $r_\lambda
=\lambda a \r(\th)$. The periods, $\tau$, the angular momenta, $L$, and
the energies,  $E$,
of these similar orbits are related by powers of the scaling parameter,
$\lambda$, as follows:
\be
\frac{\tau_\lambda}{\tau_1}=\lambda^{1+\nu/2}~,\qquad
\frac{L_\lambda}{L_1}=\lambda^{1-\nu/2}~,\qquad
\frac{E_\lambda}{E_1}=\lambda^{-\nu}~, \lb{scalr}
\ee
This scaling law can be demonstrated in various ways.
For example, Landau and
Lifshitz use
the Lagrangian to show it \cite{landau1}. We shall use
Newton's equation directly.

 For the forces corresponding to the
power potentials of Eq. \rf{potN}, we have
\be
 m \frac{d^2 {\bf r}}{dt^2}= -  \frac{d V(r)}{dr} \hat{\bf r}= -
 \frac{\nu \gamma}{r^{\nu+1}}  \hat{\bf r}~. \lb{newt}
\ee
If we choose an {\em arbitrary scale} for the distance,
\be
\r= r/a ~,
\ee
scale the time   as
\be
   {\cal T} =t/t_1~,  ~~~~~~
t_1 = a^{1+\nu/2}\sqrt{m/\gamma }~, \lb{scal1}
\ee
and  multiply Eq. \rf{newt} by  $t_1^2/(ma)= a^{\nu+1}/\gamma $,
we obtain, instead of Eq. \rf{newt}, a dimensionless equation of motion,
\bea
 \frac{d^2 {\vec{ \rho}}}{d{\cal T}^2}&=&
  \frac{t_1^2}{a} \frac{d^2 {\vec{r}}}{dt^2}= -
 \frac{ a^{\nu+1}}{\gamma} \frac{\nu \gamma}{r^{\nu+1}}
 \hat{ r}  \nonumber \\
&=&- \frac{\nu}{\r^{\nu+1}}  \hat{ \rho}~.
\eea
Therefore, the solutions of Eq. \rf{newt} must have the following form:
\be
r = r(t)= a\; \r (\th({\cal T}) ) = a\; \r(\th(t/t_1) )~,  \lb{scal}
\ee
where $\r(\th)$ describes the {\em same} functional dependence of
$\r$ on $\th$ for all similar orbits.

Thus, if we change the length scale by $a_\lambda=\lambda a$ and the
time scale according to Eq. \rf{scal1},
\be
t_\lambda
= a_\lambda^{1+\nu/2} \sqrt{m/\gamma}
=\lambda^{1+\nu/2} a^{1+\nu/2} \sqrt{m/\gamma}
=\lambda^{1+\nu/2} t_1~,
\ee
 we obtain similar solutions, which
differ in size and period. The other scaling ratios in Eq. \rf{scalr}
follow simply by dimensional analysis:
\be
\frac{L_\lambda}{L_1}=\left(\frac{a_\lambda}{a}\right)^2
\frac{t_1}{t_\lambda}= (\lambda)^2  \lambda^{-(1+\nu/2)}=
\lambda^{1-\nu/2}~,\qquad \mbox{etc.}.
\ee
The scaling property \rf{scal}  is a powerful
result.  For example,  it immediately leads to Kepler's third law for
the ratio of the periods,
$ \tau_\lambda/\tau_1= \lambda^{3/2}$, and to
the independence of the harmonic oscillator period
 on the amplitude of the motion. It also
predicts the $a-$dependence of the unit of time, $\tau_0$, that we
used in Sec. 4.4. \\

%*************************************

{\large{\bf  Appendix C: Violation of the Virial Theorem}} \\

Clearly, the violation of the virial theorem by $E=0$, $\nu > 2$
bound orbits is connected to the infinite values of the
potential and kinetic energies at $r=0$. But to better understand the
phenomenon, let us return to the  derivation
of Eq.  \rf{vir} and investigate the relevant assumptions.

One often starts  by formally differentiating the scalar product
$G={\bf p} \cdot \rb$ with
respect to time \cite{goldV}:
\be
  \dot G=\dot {\bf p} \cdot \rb + {\bf p} \cdot \dot \rb=
{\bf F} \cdot \rb + \frac{1}{m}{\bf p} \cdot {\bf p}~. \lb{virg}
\ee
Applying this result to power potentials, where
\be
  {\bf F}=-\nabla\; V(r)=- \gamma \nu r^{-\nu-1} \rbh~,
\ee
 one obtains
\be
\dot G=\nu V(r(t)) + 2 T(t)~. \lb{dotg}
\ee
Integrating Eq. \rf{dotg} with respect to time, we find
\be
G(t_2)-G(t_1)= \int_{t_1}^{t_2} \dot G= (t_2-t_1) [\nu \sp{V}+ 2\sp{T}]~,
\lb{g}
\ee
where, in  Eq. \rf{g},
 $\sp{T} $ and $\sp{V}$  again denote  time averages
of $T$ and $V$, here over the time interval $(t_1,t_2)$.

Eq. \rf{g} is valid, as long as the particle stays within one orbit
(i.e. one petal) during the  time interval.  But Eq. \rf{g}
is violated when the particle passes through the origin.
Then $G(t)={\bf p} \cdot \rb$ changes its value from $-\infty$ to
$+\infty$ at the origin,
$r=0$, meaning the derivation of the virial theorem breaks down.

To demonstrate that ${\bf p} \cdot \rb$ becomes
infinite at $r=0$, we can show that
$p$ goes to infinity stronger than $r$ goes
to zero.  From angular-momentum  conservation, we have
\be
G(t)={\bf p} \cdot \rb=rp \cos \alpha= L \cos \alpha / |\sin \alpha|=
 L \cot \alpha~,  \lb{gjump}
\ee
where $\alpha$ is the angle between
the vectors ${\bf p}$ and $\rb$. The angle $\alpha$
 changes quickly from $\pi$ to
$0$ at the $r=0$ crossing, since there $\rb$ reverses its direction from
being antiparallel to ${\bf p}$ to being parallel to ${\bf p}$. Accordingly,
$ \cot \alpha$ passes from $-\infty$ to $+\infty$.

In a similar way we can extend these results to $E<0$ orbits when
$\nu > 2 $. For clarity, we summarize our observations:\\

\noindent {\em The $E=0$ orbits of the power potentials $V=-\gamma/r^\nu$
with  $\nu \ne 2$ must either go to infinity
or pass through the origin. In addition, for $\nu>2$,
every bounded orbit with $E <0$  goes through $r=0$.}

\newpage

%****************************************************************

%*****************************************************

\newpage

\large{\bf Figure Captions}\\
\normalsize
%********************************************************************
%\baselineskip=.33in
%*********************************************************************

Figure 1:  For $\nu = 4$ a plot of the effective potential, $U$, as
a function of $r$.  (See Eq. \rf{effpot1}.)
$U(r) = [1/\r^2 - 1/\r^4]$.  \\

Figure 2:  A plot of, $P_{\nu}$, the precession per orbit, as a function of
$\nu$.  \\

Figure 3.  A plot of the dimensionless orbital period $T_{\nu} =
\tau_{\nu}/\tau_0$ as a function of $\nu$. \\

Figure 4.  The first three orbits for $\nu$ = 8.  Each orbit is precessed
$-2\pi/3$ from the previous one, so that by the end of the $3$rd orbit, the
trajectory closes.  In this and later figures, we show cartesian
coordinates for orientation.  \\

Figure 5:  The first four orbits for $\nu$ = 6.  Each orbit is precessed
$-\pi/2$ from the previous one, so that by the end of the $4$th orbit, the
trajectory closes. \\

Figure 6.  The orbit for $\nu=4$.  It is a circle, and repeats itself
continually.  \\

Figure 7:  The first two orbits for $\nu$ = 3.  Each orbit is precessed
$\pi$ from the previous one, so that by the end of the $2$nd orbit, the
trajectory closes.  \\

Figure 8:  The first two orbits for $\nu$ = 7/3.  Each orbit is precessed
$5\pi$ from the previous one, so that by the end of the $2$nd orbit, the
trajectory closes.  \\

Figure 9:  Close-up details of the beginning of the first orbit for
$\nu = 7/3$.  The orbit starts at $\th = -3\pi$.  The four graphs show the
evolution for (i) $-3\pi \leq \th \leq -2.91\pi$, (ii) $-3\pi \leq \th
\leq -5\pi/2$, (iii) $-3\pi
\leq \th \leq -2\pi$, and iv) $-3\pi \leq \th \leq -\pi$. \\

Figure 10:  The effective potential, $U(\r)=1/\r^2 - 1/\r$, as a function of
$\r$. \\

Figure 11:  A large-scale view, and a small-scale view near the origin,
of the trajectory for $\nu = 3/2$.  \\

Figure 12:  From left to right,
he trajectories for the cases i) $\nu = 1$, ii) $\nu = 0$,
iii) $\nu = -2$, and iv) $\nu = -4$.  The curves are labeled by the
numbers $\nu$.


\begin{thebibliography}{44}

\bibitem{QM1}  E. Schr\"odinger, Naturwissenschaften {\bf 14}. 664-666 (1926),
translated into English in: E. Schr\"odinger, {\it Collected Papers on Wave
Mechanics, 2nd Edition} (Blackie \& Son, London, 1928), pp. 41-44.

\bibitem{QM2}  F. Steiner, Physica B {\bf 151}, 323-326 (1988), describes
Schr\"odinger's discovery of the coherent states \cite{QM1}, and
 the interactions among he, Lorentz, Heisenberg, and others.

\bibitem{QM3}  T. T. Taylor, {\it Mechanics:  Classical and Quantum}
(Pergamon, Oxford, 1976).

\bibitem{bertrand}  M. J. Bertrand, Comptes Rendus Acad. Sci. (Paris) {\bf
77}, 849-853 (1875).

\bibitem{morse} P. M. Morse,  Phys. Rev. {\bf 34}, 57-64 (1929);

\bibitem{rm} N. Rosen and P. M. Morse, Phys. Rev. {\bf 42}, 210-217 (1932).

\bibitem{pt} G. P\"oschl and E. Teller, Zeit. Phys. {\bf 83}, 143-151 (1933).

\bibitem{morsec} W. C. DeMarcus, Am. J. Phys. {\bf 46}, 733-734 (1976).

\bibitem{rmc} M. M. Nieto and L. M. Simmons, Jr., Phys. Rev. D {\ 20},
1342-1350 (1979).

\bibitem{ptc} M. M. Nieto and L. M. Simmons, Jr., Phys. Rev. D {\ 20},
1332-1341 (1979).

\bibitem{klauder} C. Zhu and J. R. Klauder, Am. J. Phys. {\bf 61},
605-611 (1993).

\bibitem{dn2}  J. Daboul and M. M. Nieto, following paper, II

\bibitem{landau} L. D. Landau and E. M. Lifshitz, {\it Mechanics, 3rd Edition}
(Pergamon Press, N.Y., 1976).

\bibitem{landau1} Chap II, Art. 10, p. 22 in Ref. \cite{landau}.

\bibitem{gold} H. Goldstein, {\it Classical Mechanics, 2nd Edition}
(Addison-Wesley, N.Y., 1980).

\bibitem{gold1} Eq. (3.36), p. 87, in Ref. \cite{gold}.

\bibitem{problem} M. R. Spiegel, {\it Schaum's Outline of Theory and Problems
on Theoretical Mechanics} (McGraw-Hill, New York, 1967).

\bibitem{problem1}  Eq. (11),  p. 117, in Ref. \cite{problem}.

\bibitem{gold2} Fig. 3-12, p. 87, in Ref. \cite{gold}.

\bibitem{gr} I. S. Gradshteyn and I. M Ryzhik, {\it Table of Integrals,
Series, and Products} (Academic Press, New York, 1965),
integral 3.621.5.

\bibitem{bate} A. Erd\'elyi, Ed., {\it Higher Transcendental
Functions, Vol I} (McGraw-Hill, New York, 1953), p. 47, Eq. (5).

\bibitem{problem3}  Problem 5.60,  p. 138, in Ref. \cite{problem}.

\bibitem{problem2}  Sec. 5.15,  p. 127, in Ref. \cite{problem}.

\bibitem{marion} J. B. Marion and  S. T. Thornton, {\it Classical Dynamics,
3rd Edition} (Harcourt Brace Jovanovich, Publishers, N.Y., 1988), p. 287.

\bibitem{cotes} Chap. IV, Art. 15, problem 2, p. 40 in Ref. \cite{landau}.

\bibitem{moregold}  Sec. 9-7, p. 420, in Ref. \cite{gold}.

\bibitem{schiff} L. \ I. \ Schiff, {\it Quantum Mechanics, 3rd Edition}
(McGraw-Hill, New York, 1968), Ch. 7, Sec. 30, p. 234.

\bibitem{disc} The discontinuity at the origin of the azimuthal
angle, $\th$, by $\pi$,
 is only
 a technical problem associated with the nature
of polar coordinates.  It  disappears in cartesian
coordinates because both $x(t)$ and $y(t)$ are
continuous functions of time.

\bibitem{goldV} Sec.  3-4, p. 82 in Ref. \cite{gold}.

\bibitem{landauV} Chap. II, Art. 10, p. 23 in Ref. \cite{landau}.

\end{thebibliography}
\end{document}